\documentclass[reprint,amsmath,amssymb,aps,prc]{revtex4-2}
\pdfoutput=1
\usepackage{graphicx} 
\usepackage{dcolumn} 
\usepackage{bm} 
\usepackage{subcaption}
\usepackage{caption}
\usepackage{float}
\usepackage{hyperref}
\usepackage{multirow}
\usepackage{xcolor}
\hypersetup{colorlinks=true,linkcolor={blue},citecolor={red},urlcolor={blue}}
\begin{document}
\title{On the thermalisation of leptons from heavy-flavour meson decay in pp collisions}
\author{Rahul Ramachandran Nair}
\email{physicsmailofrahulnair@gmail.com}
\thanks{\\Orcid ID: 0000-0001-8326-9846}
\affiliation{National Centre For Nuclear Research, 02-093 Warsaw, Poland}
\altaffiliation{The work was performed while the author was affiliated with the National Centre For Nuclear Research, Warsaw, Poland; Present address: Sreeraj Bhavan, Karukachal PO, Kottayam(District), Kerala, India- 686540\\} 
\date{\today}
\begin{abstract}
It is shown that the $\mu^{\pm}$ originating from the decay of heavy-flavour mesons in the simulated proton-proton collisions at $\sqrt{s} = 13$ TeV using the PYTHIA 8 event generator can be made into two groups based on the value of their light front variable such that one among these two groups of $\mu^{\pm}$ follow the Boltzmann statistics with a nearly isotropic angular distribution. A similar effect is not observed for $e^{\pm}$ from the decay of heavy-flavour mesons in these collisions.
\end{abstract}
\maketitle
\begin{table*}[hbt!]
\begin{center}
\begin{tabular}{ c  c c c c c c c } 
\noalign{\smallskip}\hline\noalign{\smallskip}
$\sqrt{s}$ (TeV) & $\zeta_c$ & $T_{\zeta}$(MeV)& $\chi^2/ndf$&  $T_{p_T^2}$(MeV) & $\chi^2/ndf$  & $T_{cos(\theta)}$ (MeV)& $\chi^2/ndf$\\ 
\noalign{\smallskip}\hline\noalign{\smallskip}  
$13.0$ & 10.30 & 117 $\pm$ 5  & 24/26 &  80 $\pm$ 3 & 28/26  & 94 $\pm$ 7 & 33/23\\  
\noalign{\smallskip}\hline\noalign{\smallskip} 
\end{tabular}
\end{center}
\caption{Temperatures obtained from the light front analysis of $\mu^{\pm}$ from heavy flavour meson decay in the simulated proton-proton  collisions at $\sqrt{s}= 13$ TeV with PYTHIA 8.}
\label{Table}
\end{table*}
The charm and beauty quarks have their masses larger than the non-perturbative scale of QCD ($m_q > \Lambda_{QCD}$) and the typical temperature of deconfined QCD matter ($m_q > T_{QGP}$). They are formed during the initial hard scattering stage of ultrarelativistic heavy-ion collisions with the thermalisation times comparable to the lifetime of supposedly formed fireball in such collisions \cite{doi:10.1146}. Hence, the heavy quarks are expected to witness the full evolution of the system and are considered to be unique probes to the hadronisation \cite{Greco_2004} and dynamics originating from a quark-gluon plasma-like medium in such collisions \cite{THOMA1991491, GYULASSY1990432}. The production cross-section of these 'heavy' quarks can be calculated using the perturbative QCD approach \cite{PhysRevLett.95.122001}. The heavy flavoured hadrons formed from the hadronisation of heavy quarks may be measured through their non-leptonic weak decays or semi-leptonic decay while the experimental estimation at the collider experiments is a challenging task with the backgrounds coming from multiple sources \cite{LAPOINTE2014213}. The specific dynamics of heavy-ion collisions are not expected to be present in the pp collisions. Hence, the measurement of leptons from heavy-flavour hadron decays in the proton-proton (pp) collisions are considered as a benchmark to interpret the results from heavy-ion collisions. Though this remains the case, concerns arise on this bench-marking since the typical effects in heavy-ion collisions such as collectivity and strangeness enhancement were observed in high multiplicity pp collisions at LHC \cite{Adam2017, Khachatryan_2017}. In this paper, we explore the thermalisation of leptons from heavy flavour meson decay (D and B mesons) in the pp collisions at LHC energies in the framework of 'light front analysis' with PYTHIA 8 event generator \cite{Sj_strand_2006, Sjostrand:2007gs}.\\\\
The light front analysis scheme was originally proposed in \cite{Garsevanishvili78, Garsevanishvili79} for studying the production process of charged pions in hadron-hadron and nucleus-nucleus interactions. In the centre of mass frame, the light front variables can be written in the following scale invariant form:
\begin{equation} \label{EqnZeta}
\xi^{\pm} = \pm \frac{E + |p_{z}|}{\sqrt{s}} 
\end{equation}
or conveniently 
\begin{equation} \label{EqnZeta}
\zeta^{\pm} = \mp \ln \left(\xi^{\pm}\right)
\end{equation}
where, $s$ is the Mandelstam variable, $p_{z}$ is the z-component of the momentum and E is the energy of the particle. The positive and negative signs in Eq. \eqref{EqnZeta} corresponds to the positive and negative hemispheres. A pronounced maximum in the $\zeta^{\pm}$ distribution of inclusively produced charged pions was observed around a light front value denoted by $\tilde{\zeta}^{\pm}$ in the low energy experiments \cite{Amaglobeli99, Djobava03}. The particles with their $\zeta^{\pm} > \tilde{\zeta}^{\pm}$ were observed to have a comparatively isotropic polar angle distribution with respect to those particles having  $\zeta^{\pm} < \tilde{\zeta}^{\pm}$. The $p_{T}^{2}$ distribution for these two groups of particles were found to have different slopes as well. From these observations, it was hypothesised that a thermalisation has been reached in the region of the phase space of the charged pions with $\zeta^{\pm} > \tilde{\zeta}^{\pm}$. Note that a constant value of the $\zeta^{\pm}$ defines a paraboloid in the phase space given by
\begin{equation}
p_{z}  = \frac{p_T^2+m^2- (\tilde{\xi}^{\pm}\sqrt{s})^2}{-2\tilde{\xi}^{\pm}\sqrt{s}}
\label{paraboloid}
\end{equation}
To investigate the hypothesis of thermalisation of charged pions with  $\zeta^{\pm} > \tilde{\zeta}^{\pm}$ a set of analytic expressions were constructed and tested against the experimental data using a simple statistical model in \cite{Amaglobeli99, Djobava03}. A similar analysis is presented here with the $e^{\pm}$ and $\mu^{\pm}$ from the decay of heavy-flavour mesons in the simulated pp collisions at $\sqrt{s} = 13$ TeV using the perturbative QCD based PYTHIA (Version 8.303) event generator \cite{Sj_strand_2006, Sjostrand:2007gs} to investigate their thermal equilibration. 50 million events with at least 5 charges particles in the mid-rapidity region are simulated with Monash 2013 tune of PYTHIA \cite{Skands_2014}. The inelastic, non-diffractive component of the total cross-section is implemented for all soft QCD processes. The multi partonic interaction based scheme of colour re-connection mechanism in PYTHIA 8 is also implemented. The  $e^{\pm}$ and $\mu^{\pm}$ produced from the decay of heavy-flavour mesons in these collisions are selected and a distribution of $|\zeta^{\pm}|$ is made for each of them. The $|\zeta^{\pm}|$ distributions are fitted with the following equation
\begin{equation}
\frac{dN}{d\zeta^{\pm}} \sim \int_0^{p_T^2(max)} E f(E)dp_T^2
\label{ZetaInt}
\end{equation}
where $p_{T,max}^2$ is given by
\begin{equation}
p_{T,max}^2 = (\xi^{\pm}\sqrt{s})^{2} - m^{2} \label{ptmax}
\end{equation}
The lowest value of $|\zeta^{\pm}|$ to which it can be done successfully is taken as $|\tilde{\zeta^{\pm}}|$. We use the $\chi^2$ minimisation method for fitting as implemented in the ROOT (version 6.18/04) software package \cite{BRUN199781}. For a successful $\chi^2/n.d.f \sim 1.0$ where $n.d.f$ stands for the number of degrees of freedom. The particles are then divided into two groups based on the value of the $|\tilde{\zeta^{\pm}}|$. The $|cos(\theta)|$ distribution of particles with $|\zeta^{\pm}| > |\tilde{\zeta^{\pm}}|$ is made and is fitted with following equation 
\begin{equation}
\frac{dN}{dcos(\theta)} \sim \int_0^{p_{max}} f(E) p^2dp
\label{CosInt}
\end{equation}
where $p_{max}$ is given by
\begin{equation}
p_{max}  = \frac{-\tilde{\xi^{\pm}}\sqrt{s}cos(\theta) + \sqrt{(\tilde{\xi^{\pm}}\sqrt{s})^2  - m^2 sin^2(\theta)}}{sin^2(\theta)}
\label{pmax}
\end{equation}
\begin{figure}[hbt!] 
\begin{subfigure}{0.50\textwidth}
\includegraphics[width=\linewidth]{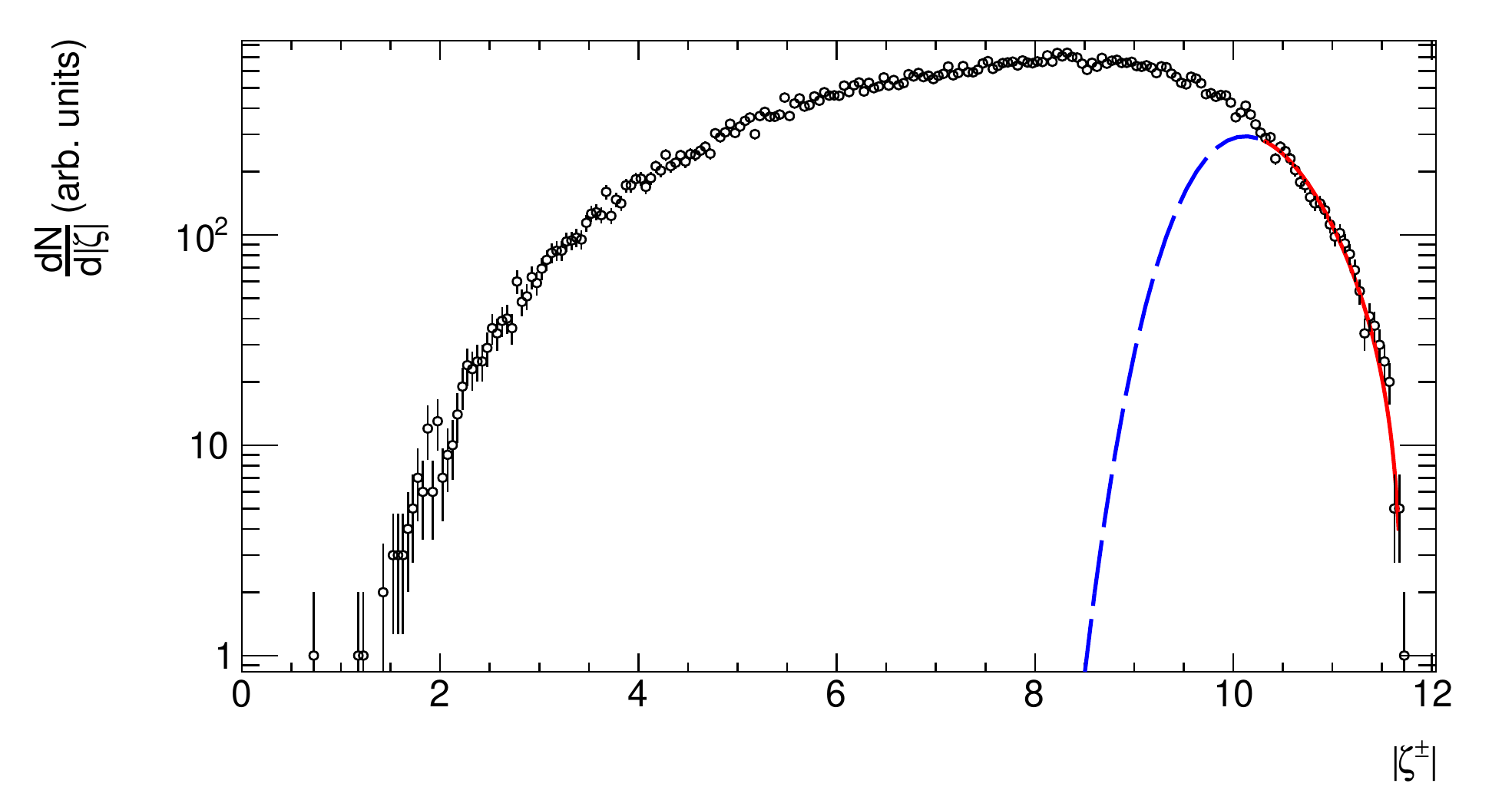}
\caption{$|\zeta^{\pm}|$ distribution of $\mu^{\pm}$; $\zeta_c = 9.80$ }\label{ZetaMuPP}
\end{subfigure}\\
\begin{subfigure}{0.50\textwidth}
\includegraphics[width=\linewidth]{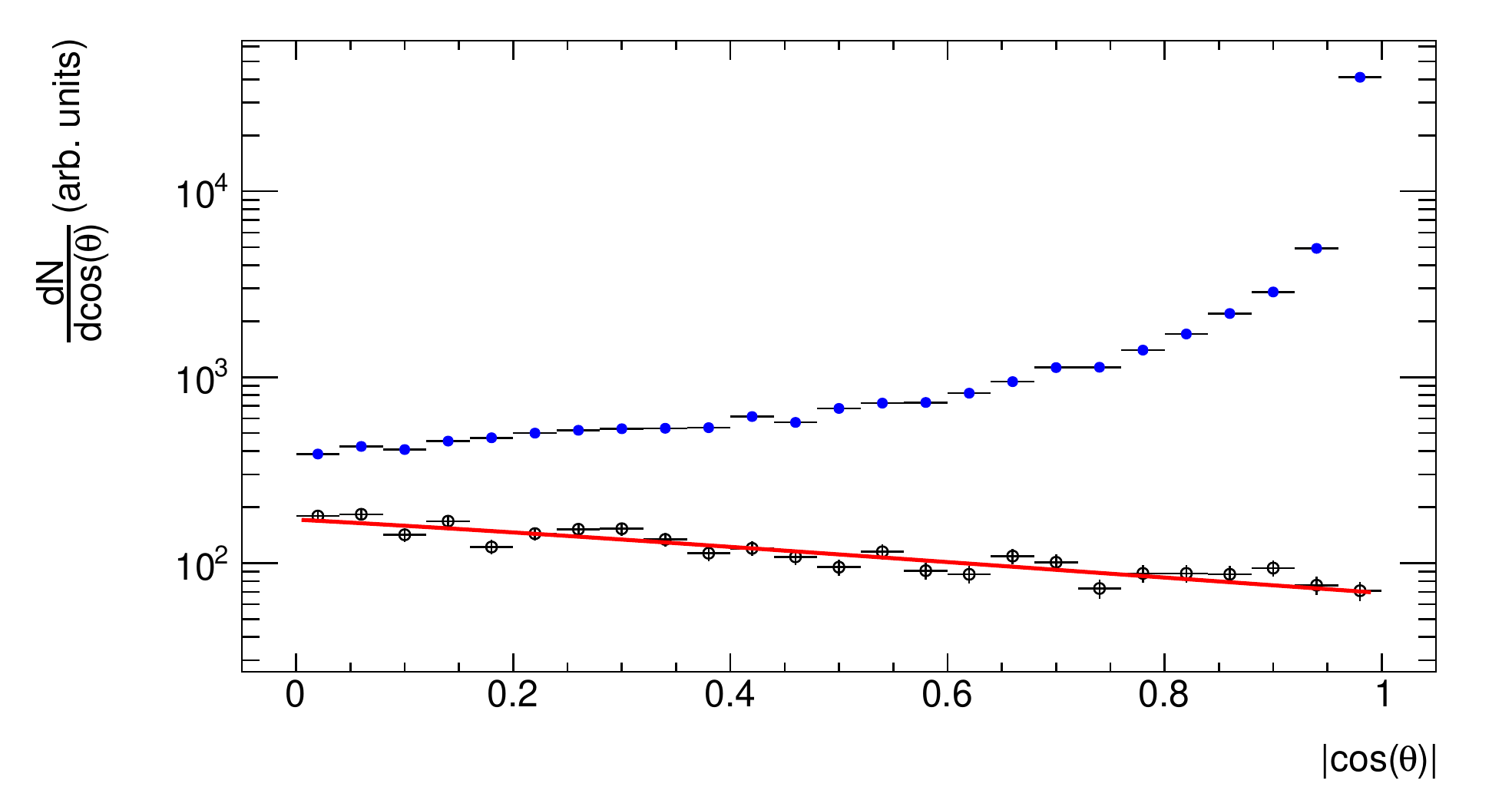}
\caption{$|cos(\theta)|$ distribution of $\mu^{\pm}$; $\zeta_c = 10.30$}
\label{CosMuPP}
\end{subfigure}\\
\begin{subfigure}{0.50\textwidth}
\includegraphics[width=\linewidth]{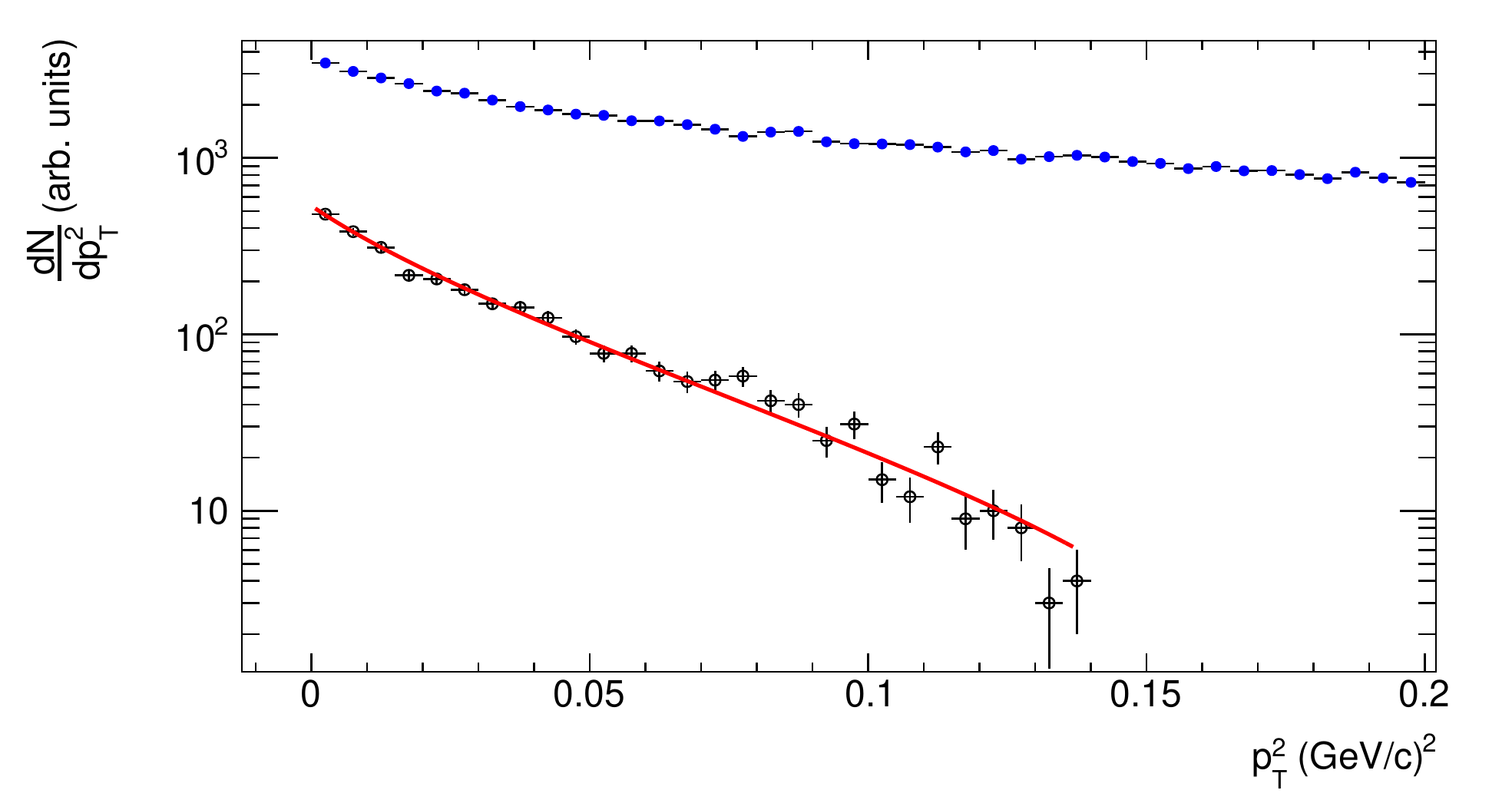}
\caption{$p_T^2$ distribution of $\mu^{\pm}$; $\zeta_c = 10.30$}
\label{PtSqMuPP}
\end{subfigure}
\caption{Marked in open circles are the distribution of $\mu^{\pm}$ from heavy flavour meson decay with $\zeta > 10.30$ and solid red curves are the results of the fit. The distributions marked in blue are of those $\mu^{\pm}$ from heavy flavour meson decay with $|\zeta^{\pm}| < 10.30$ } \label{Mu-pp}
\end{figure}
\begin{figure}[hbt!] 
\begin{subfigure}{0.50\textwidth}
\includegraphics[width=\linewidth]{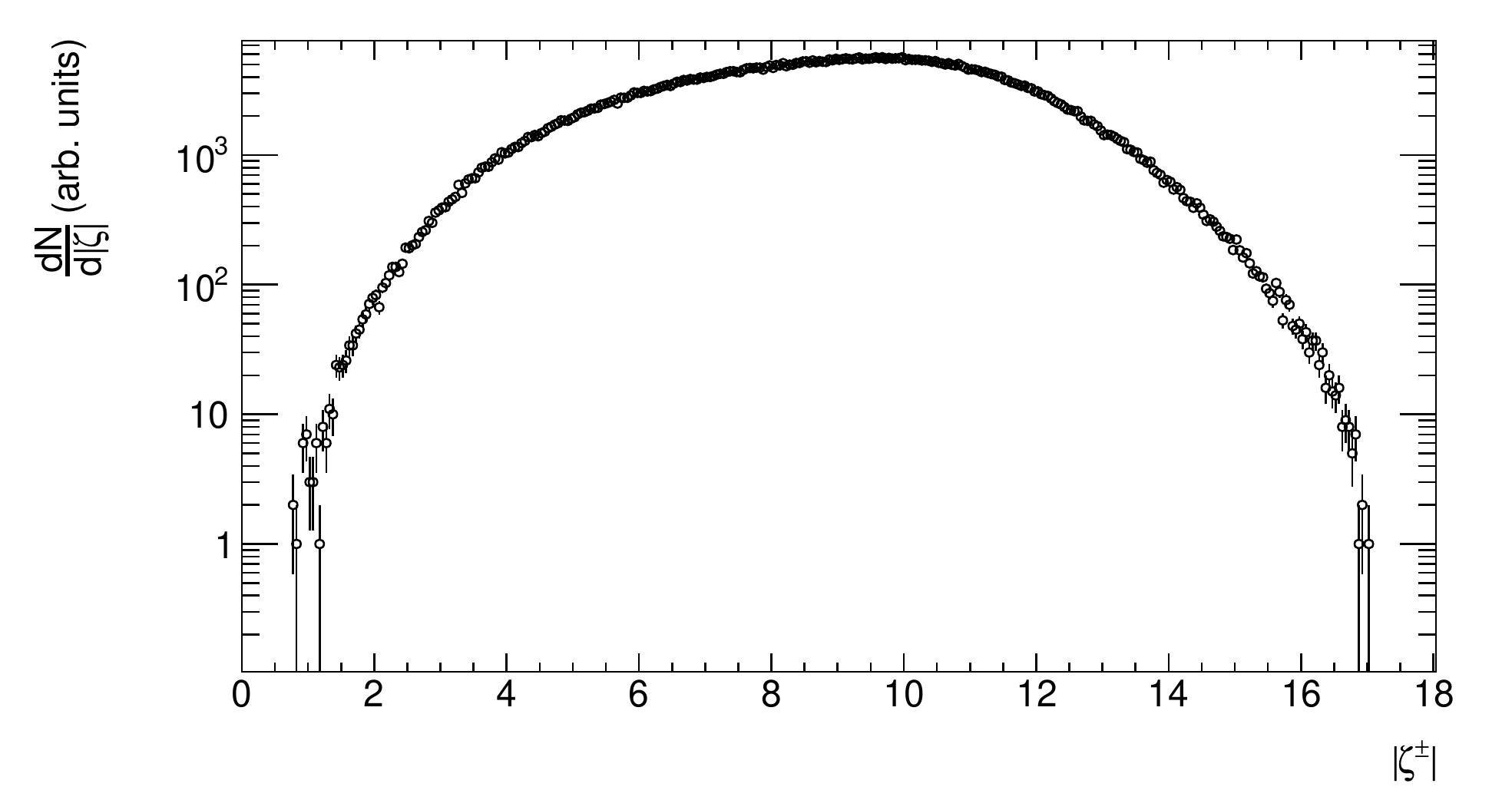}
\caption{$|\zeta^{\pm}|$ distribution of $e^{\pm}$}\label{ZetaE}
\end{subfigure}\\
\begin{subfigure}{0.50\textwidth}
\includegraphics[width=\linewidth]{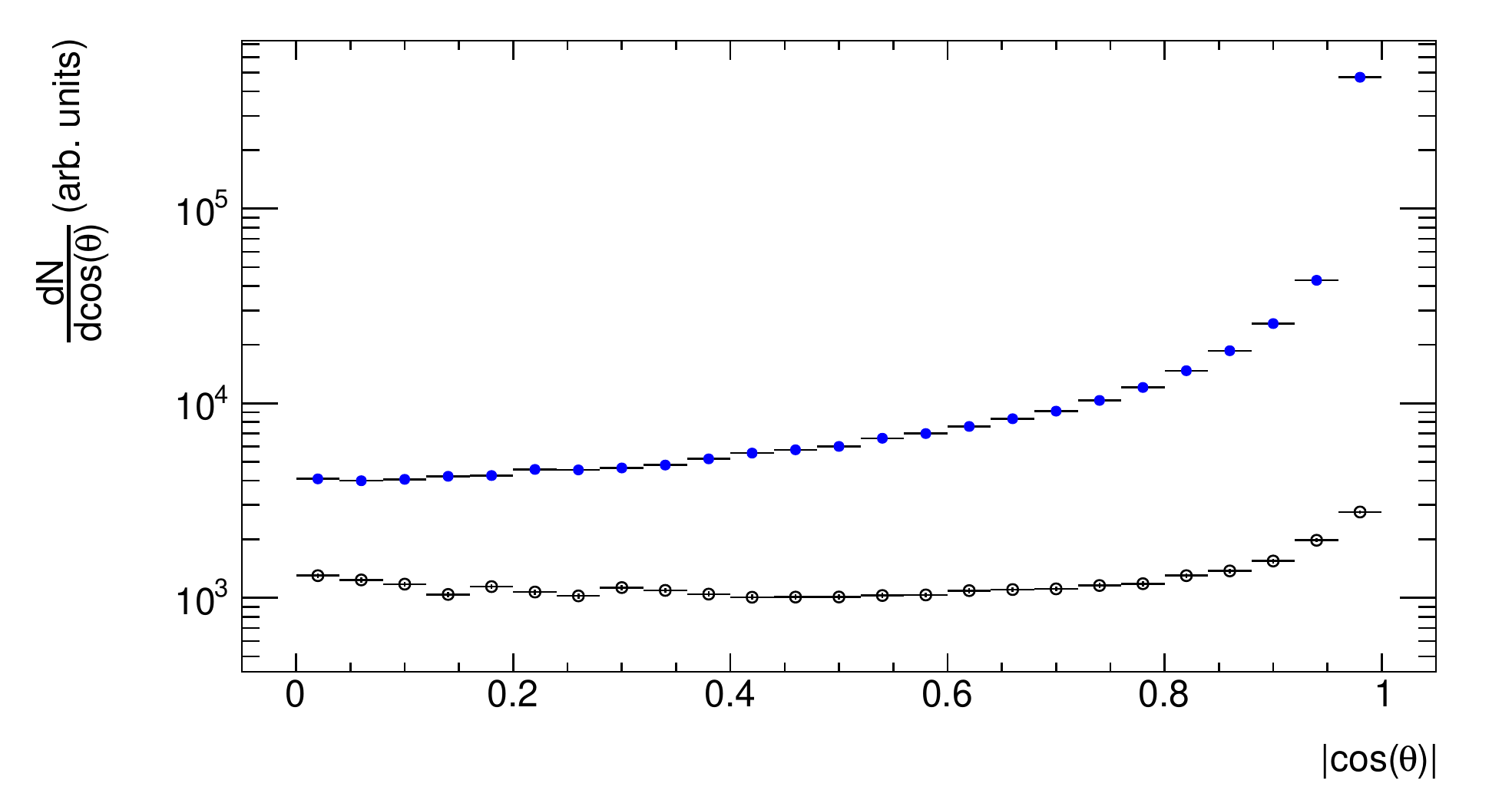}
\caption{Marked in open circles are the distribution of $e^{\pm}$ with $\zeta > 13.0$ and marked in blue are of those $e^{\pm}$ with $\zeta < 13.0$}
\label{CosE}
\end{subfigure}
\caption{Distributions of $e^{\pm}$ from heavy flavour meson decay in pp collisions at $\sqrt{s} = 13.0$ TeV} \label{E-pp}
\end{figure}
The $|\tilde{\zeta^{\pm}}|$ in the limit of the integral in Eq. \eqref{CosInt} denoted by $p_{max}$, is the value of $|\tilde{\zeta^{\pm}}|$ obtained from fitting the $|\zeta^{\pm}|$ distribution. The $p_{T}^{2}$ distribution of particles with $|\zeta^{\pm}| > |\tilde{\zeta^{\pm}}|$ is made and fitted with equation 
\begin{equation}
\frac{dN}{dp_T^2} \sim \int_0^{p_{z,max}} f(E)dp_{z}
\label{PtSqInt}
\end{equation}
using the same value $|\tilde{\zeta^{\pm}}|$ for calculating $p_{z,max}$. The expression for $p_{z,max}$ is given by
\begin{equation}
p_{z,max}  = \frac{m^2 + p_T^2 - (\tilde{\xi^{\pm}}\sqrt{s})^2}{-2\tilde{\xi^{\pm}}\sqrt{s}}
\label{pzmax}
\end{equation}
In all these three cases of the fitting, the form of $f(E)$ is of the Boltzmannian form given by
\begin{equation}\label{eqboltz}
f(E) \sim  \exp(-E/T)
\end{equation} 
If the three fits are successful, then $|\tilde{\zeta^{\pm}}|$ is taken as the final value $\zeta_c$ of the light front variable. If any of these three fits is not successful, we repeat the procedure with a larger value of $|\tilde{\zeta^{\pm}}|$ until the three fits are successful or the value of $|\zeta^{\pm}|$ can no longer be increased.
By picking $\zeta_c$ as 10.30 for the $\mu^{\pm}$ from heavy-flavour meson decay in the simulated pp collisions, we could find a group of particles following the Boltzmann statistics. The results of the fits are shown in FIG. \ref{Mu-pp}. The temperatures obtained for each case are summarised in TABLE \ref{Table}. However, the light front analysis could not find a group of thermalised $e^{\pm}$ from heavy-flavour meson decay for any value of its $\zeta_c$. The $\zeta$ distribution $e^{\pm}$ from heavy-flavour meson decay is shown FIG.\ref{ZetaE} and the polar angle distribution of those $e^{\pm}$ with $\zeta > 13.0$ and $\zeta < 13.0$ are shown in FIG.\ref{CosE}. A nearly isotropic polar angle distribution was found to be missing for $e^{\pm}$ for all the possible values of $\zeta_c$ we tried with. The $p_{T}^{2}$ values of $e^{\pm}$ with $\zeta > 13.0$ are extremely small ($p_{T}^{2} < 0.005$).\\\\
Thus $e^{\pm}$ from the decay of heavy-flavour mesons do not appear to be thermalised while a nearly isotropic set of $\mu^{\pm}$ following Boltzmann statistics could be found out with the light front analysis. The reason for such an effect is unknown and might be interesting to explore in the light of the observed lepton-flavour dependence. In the light front analysis of inclusive hadrons in pp collisions at LHC energy using PYTHIA 8 \cite{nair2021light} and Au-Au collisions at RHIC energy using the UrQMD model \cite{nair2021polynomial}, it was demonstrated that the $p_T$ dependence of $\zeta_c$ can be quantified to select a group of particles following the Boltzmann statistics. It would be interesting to perform such an analysis for the  $e^{\pm}$ and $\mu^{\pm}$ from heavy-flavour decay with the experimental data at LHC. Since the PYTHIA 8 model does not explicitly incorporate a QGP-like medium in it, a similar light front analysis with the experimental data at LHC might reflect the dynamics of QGP-like medium in the experimental collisions upon comparison with the results of our kind of analysis in the proper acceptance region of the relevant detectors.\\\\
\textit{I thank Harsh Shah and Smita Chakraborty (Lund University) for the kind suggestions and hints given in the context of the heavy-ion collision module in Pythia.}

\bibliography{article}
\end{document}